\documentclass[12pt]{revtex4}
\usepackage{epsfig}
\usepackage{bbm}

\begin{document}

\rightline{ITP-UU-04/26, SPIN-04/14}

\title{Photon mass in inflation and \\ nearly minimal magnetogenesis
\footnote{Contribution to the Proceedings of Strong and Electroweak Matter 2004
(SEWM2004), Helsinki, Finland, June 16-19, 2004.}}

\author{Tomislav Prokopec}

\address{
        Institute for Theoretical Physics \& Spinoza Institute,
        Leuvenlaan 4, 3584 CE Utrecht, The Netherlands;
        {\tt Email: \hskip -0.21cm T.Prokopec@phys.uu.nl} 
        }

\maketitle


\centerline{\bf Abstract}

 We argue that the dynamics of photons canonically coupled to 
           scalars in de Sitter inflation gets modified by a photon mass term.
         This might have lead to generation of potentially observable magnetic
          field correlated over cosmological scales today.

\section{Introduction}
\label{introduction}

 One of the important unresolved puzzles of modern astrophysics
and cosmology is the origin of large scale magnetic fields of the Universe. 
It is conceivable that galactic magnetic fields are explained
by the dynamo mechanism, by which an initial small seed field
 at the time of galaxy formation was amplified to the microgauss strength
observed today. Seed fields of magnitudes $B_{\rm seed} \sim 10^{-20}$~Gauss
and correlated over distances larger than about 100 parsecs
could have been created by the Biermann battery mechanism
induced by the large scale structure dynamics at the time of galaxy formation. 
On the other hand, it seems that no currently available astrophysical 
scenario can explain recent cluster fields~\cite{VogtEnsslin:2003} observed
in several galaxy clusters. 

 Based on a study of scalar electrodynamics ($\Phi$ED)
 during inflation
\cite{ProkopecPuchwein:2004,ProkopecPuchwein:2003,ProkopecWoodard:2003photons,ProkopecTornkvistWoodard:2002AoP,ProkopecTornkvistWoodard:2002prl,ProkopecWoodard:2003ajp,DimopoulosProkopecTornkvistDavis:2001,DavisDimopoulosProkopecTornkvist:2000}
 and subsequent radiation and matter epochs, in this talk I will argue
 that the observed large scale magnetic fields could have originated
in cosmological inflation.

\section{The Model}
\label{The Model}

Cosmological inflation is an epoch of a rapid (accelerated) expansion of the 
Universe. For simplicity, here we model inflation by de Sitter space-time.
In conformal coordinates it is characterized by the metric,
$g_{\mu\nu} = a^2 \eta_{\mu\nu}$, where $\eta_{\mu\nu} = {\rm diag}(-1,1,1,1)$
and the scale factor $a= - 1/(H\eta)$, $H$ is the Hubble parameter and 
$\eta$ conformal time. In conformal coordinates the Lagrangian of scalar
electrodynamics reduces to 
\begin{equation}
 {\bf L}_{\rm\Phi ED}
          = - \frac{1}{4}\eta^{\mu\nu}\eta^{\rho\sigma}F_{\mu\rho}F_{\nu\sigma}
            - a^2\eta^{\mu\nu}(D_\mu \Phi)^*(D_\nu \Phi) 
            - a^4(m_\phi^2 +\xi{\rm R})\Phi^*\Phi
\,, 
\label{calL} 
\end{equation}
where $F_{\mu\rho}=\partial_\mu A_\rho-\partial_\rho A_\mu$ is the 
photon field strength, $D_\mu = \partial_\mu - ieA_\mu$ is the covariant 
derivative, $m_\phi$ the scalar field mass, ${\rm R}$ the curvature scalar, 
which in $D=4$ and de Sitter inflation reads ${\rm R} = 12 H^2$, and 
$A_\mu$ is the gauge (photon) field. When $m_\phi^2 + \xi \rm R \neq 0$,
conformal invariance of $\Phi$ is broken during inflation, resulting in 
a growth of scalar fluctuations primarily on superhorizon scales.  
This is formally encoded by the 
Chernikov-Tagirov~\cite{ChernikovTagirov:1968} (Bunch-Davies) 
Feynman propagator, which in $D=4$ reads
\begin{equation}
   iG_F(y) = \frac{H^2}{4\pi^2}\Gamma\big(\frac 32 -\mu\big)
                               \Gamma\big(\frac 32 +\mu\big)
   \; {}_2F_1\Big(\frac32-\mu,\frac32+\mu,2;1-\frac{y}{4}\Big)
\,,
\label{Chernikov-Tagirov}
\end{equation}
where $\mu = [(3/2)^2 - (m_\phi^2 + \xi {\rm R})/H^2]^{1/2}$,
and $y = a(\eta)a(\eta^\prime)\Delta x^2$ is the de Sitter invariant length,
$\Delta x^2 = - (|\eta-\eta^\prime|-i\epsilon)^2 + \|\vec x-\vec x^\prime\|^2$.
When expanded in $s \equiv (3/2) - \mu \approx (m_\phi^2/3H^2) + 4\xi$
($|s|\ll 1$), the propagator simplifies to~\cite{ProkopecPuchwein:2003} 
\begin{equation}
   iG_F(y) \simeq  \frac{H^2}{4\pi^2}\Big\{\frac 1y 
                                         - \frac 12 \ln(y)
                                         +\frac{1}{2s} - 1 + \ln(2)
                                     \Big\}
\,.
\label{Chernikov-Tagirov:2}
\end{equation}
Note that this propagator is still fully de Sitter invariant. 
In the limit when $m_\phi\rightarrow 0$ and $\xi \rightarrow 0$ however,
the propagator diverges~(\ref{Chernikov-Tagirov}), and
a natural choice is an $SO(3,1)$ invariant 
propagator~\cite{AllenFolacci:1987,ProkopecPuchwein:2003},
\begin{equation}
 iG_F|_{m_\phi=0} \simeq \frac{H^2}{4\pi^2}\Big\{\frac 1y 
                                   - \frac 12 \ln(y) 
                                   + \frac12 \ln\big(a(\eta)a(\eta^\prime)\big)
                                   - \frac 14 + \ln(2)
                                 \Big\}
\,.
\label{Allen-Folacci}
\end{equation}
The de Sitter invariance $SO(4,1)$ is broken by the term 
$\propto \ln\big(a(\eta)a(\eta^\prime)\big)$.
We shall use both propagators~(\ref{Chernikov-Tagirov:2}) 
and~(\ref{Allen-Folacci}) to study how scalars influence the photon dynamics
during inflation and its consequences today.

 During inflation the photon dynamics in $\Phi$ED~(\ref{calL})
is governed by the modified Maxwell equation,
\begin{equation} 
 \eta^{\mu\nu}\partial_\mu F_{\nu\rho}
    + \eta_{\rho\mu}\int d^4x^\prime [{}^\mu\Pi^\nu_{\rm ret}](x,x^\prime) 
               A_{\nu}(x^\prime) = 0
\,,
\label{eom}
\end{equation}
where $[{}^\mu\Pi^\nu_{\rm ret}]$ denotes the retarded 
vacuum polarization tensor of the theory. We have 
calculated~\cite{ProkopecPuchwein:2003,ProkopecTornkvistWoodard:2002AoP} 
the one-loop photon vacuum polarization tensor by making use of the 
Schwinger-Keldysh formalism, suitable for study of time dependent problems. 
Our polarization tensor is renormalized 
by the technique of dimensional regularization. Further, it is
manifestly transverse, and hence gauge invariant.
We have performed both calculations, for photons coupled to a massive light 
scalar~(\ref{Chernikov-Tagirov:2})~\cite{ProkopecPuchwein:2003}, 
as well as to a massless minimally
coupled scalar~(\ref{Allen-Folacci})~\cite{ProkopecTornkvistWoodard:2002AoP}. 
In both cases the photon acquires a mass through the coupling to the infrared 
scalar modes. The result for the case of coupling to massive excitations
 can be neatly summarized by the effective
 Lagrangian~\cite{ProkopecPuchwein:2003}
\begin{equation}
 {\bf L}_{\rm\Phi ED}^{\rm eff} =  {\bf L}_{\rm\Phi ED}
                   - \frac 12 a^2 \eta^{\mu\nu} A_{\mu}A_{\nu}
\,,
\label{L:effective}
\end{equation}
such that the photon obeys a Proca equation of motion, with the mass
\begin{equation}
  m_\gamma^2 \simeq \frac{3\alpha H^4}{\pi(m_\phi^2 + \xi {\rm R})}
\,,
\label{photon-mass}
\end{equation}
where $\alpha = e^2/(4\pi)$ denotes the fine structure constant. 
This result is obtained in a generalized Loren(t)z gauge, 
$\eta^{\mu\nu}\partial_\mu(a^2 A_\nu) = 0$. 
Both transverse and longitudinal vector excitations are endowed by
the mass~(\ref{photon-mass}). This represents an alternative, 
gravity induced, mechanism for gauge field mass generation, in which 
the longitudinal photon degree of freedom is provided by the infrared 
modes of the (complex) scalar field. 
As a consequence of mass generation, scaling of the electric and magnetic
fields during inflation changes from the vacuum scaling,
$\vec E_{\rm vac}, \vec B_{\rm vac} \propto a^{-2}$ to 
$\vec E \propto a^{ - \frac32 - \frac 12[1-4(m_\gamma/H)^2]^{1/2}}$
and
$\vec B \propto a^{ - \frac52 + \frac 12[1-4(m_\gamma/H)^2]^{1/2}}$.
If the photon couples to a massless minimally coupled scalar, 
then the magnetic field remains massless, while the electric field acquires
a time dependent mass, 
$m_\gamma^2(a) = (2\alpha H^2/\pi)[\ln(a)-1/4]$,
\cite{ProkopecWoodard:2003photons}
such that the fields scale differently, $\vec E\propto a^{-3/2}$
 and $\vec B\sim \vec B_{\rm vac} \propto a^{-2}$.

\section{Nearly minimal magnetogenesis}
\label{Nearly minimal magnetogenesis}

 We shall now argue that, provided the scalar field evolution remains
de Sitter invariant~(\ref{Chernikov-Tagirov:2}), 
a certain choice of the couplings leads to magnetic field generation
during inflation. 

A general solution of the photon mode equation associated with
the Proca effective theory~(\ref{L:effective}) can be written
in terms of the Hankel function, 
\begin{equation}
     A_\mu = \frac 12 \sqrt{-\pi\eta}H_{\nu}^{(1)}
                    (-k\eta)\varepsilon_\mu^{(1)}
\,,\qquad
         \nu = \frac 12\big(1 - 4{m_\gamma^2}/{H^2}\big)^{1/2}
\label{A:inflation}
\end{equation}
\hskip -0.15cm
where $\varepsilon_\mu^{(1)}$ denotes the photon polarization vector.
This needs to be matched to the plane wave solutions of the 
radiation era, $A_\mu^{\rm (rad)} = (2k)^{-1/2}{\rm e}^{\mp ik\eta}$.
These modes describe correctly the photon dynamics in radiation era,
provided conductivity is small and the photon mass
decays nonadiabatically during radiation era.
After performing the mode matching, we arrive at the following 
ensemble averaged magnetic field in radiation era,
\begin{equation}
 \langle \vec B^2(\vec x,\eta,\ell)\rangle = \int \frac{dk}{k}
            {\bf P}_B(k,\eta,\ell) |w(k,\ell)|^2
\,,
\label{photon:spectrum}
\end{equation}
where $w$ denotes a window function, {\it e.g.}
$w = \exp(-k^2\ell^2/2) \equiv \int d^3xW(\vec x)\exp(i\vec k\cdot\vec x)$,
$W = (2\pi\ell^2)^{-3/2}\exp\big(-\|\vec x\|/(2\ell^2)\big)$,
and ${\bf P}_B(k,\eta,\ell)$ is the field spectrum 
(defined in analogous manner as the spectrum of cosmological perturbations),
\begin{equation}
 {\bf P}_B = \frac{1}{a^4}\frac{\Gamma^2(\nu)}{8\pi^3}
            \big(\nu-\frac12\big)^2(2H)^{2\nu+1}
             k^{3-2\nu}\sin^2\big((k/H)(a-1)\big)
\,.
\label{PB:spectrum}
\end{equation}
\vskip -0.20in
\begin{figure}[htbp]
\centerline{\epsfxsize=3.in\epsfbox{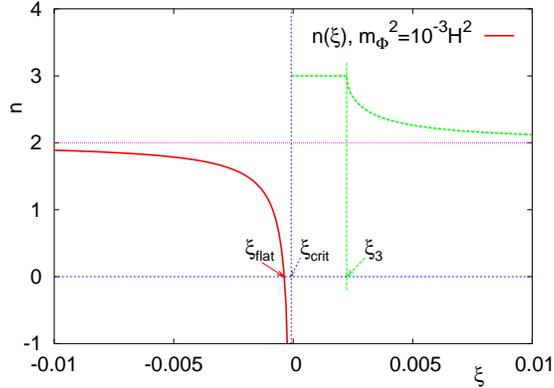}}   
\caption{Spectral index $n$ as a function of 
the scalar field coupling to curvature $\xi$ and the scalar mass $m_\phi^2$.}
\label{fig:one}
\end{figure}
This spectrum is characterized by the spectral index $n = 3-2\nu$
 on subhorizon scales, and by $n = 5-2\nu$ on superhorizon scales. 
Figure~\ref{fig:one} shows $n$ on subhorizon scales as a function of 
 $\xi$ and
 $m_\phi^2$.
Note that for $\xi_{\rm flat} = - (\alpha/8\pi) - m_\phi^2/12H^2$,
one gets a flat spectrum, $n=0$. In this case 
the field strength is typically $B\sim 10^{-10}$~Gauss on all scales,
such that it is potentially observable by the next generation of 
CMBR measurements. 
The region of couplings $\xi \in (\xi_{\rm flat},\xi_{\rm crit})$
$\;(\xi_{\rm crit} = -m_\phi^2/12H^2)$
leads to a growth in magnetic energy during inflation, such that 
its backreaction on the background space-time may become important.
This may change the course of the Universe's evolution, and thus
$n<0$ cannot be trusted. The values corresponding to
$\xi > \xi_3 = (\alpha/\pi) - m_\phi^2/12H^2$ have already been 
proposed
earlier~\cite{DavisDimopoulosProkopecTornkvist:2000,DimopoulosProkopecTornkvistDavis:2001,ProkopecWoodard:2003ajp} 
as a model for magnetic field generation,
powerful enough to seed the galactic dynamo.

 Finally, we recall that radiation era may be endowed by a large conductivity.
In this case the photon dynamics is governed by the suitably 
modified~\cite{ProkopecPuchwein:2004} B\"odeker-Langevin 
theory~\cite{Bodeker:1999},
\begin{equation} 
(a \sigma \partial_t + \vec k^2 )A^T \!= a^3 \zeta^T
\!,\;
 \langle\zeta^T\!(\vec k,\eta)\zeta^{T^\prime}\!(\vec k^\prime,\eta)\rangle
         \!=\!\frac{2\sigma T}{a^4}\delta^{TT^\prime}\!\!
              (2\pi)^3\delta(\eta-\eta^\prime)
                \delta(\vec k - \vec k^\prime)
\label{BodekerLangevin}
\end{equation}
\vskip -0.2cm
\noindent
and the spectrum~(\ref{PB:spectrum}) is modified to 
\begin{equation}
 {\bf P}_B = \frac{1}{a^4}\frac{\Gamma^2(\nu)}{2\pi^3}
            (2H)^{2\nu-1}
             k^{5-2\nu}\exp\Big(-\frac{2k^2}{H\sigma}(1-1/a)\Big)
            +  {\bf P}_B^{\rm th}
\,.
\label{PB:spectrum:sigma}
\end{equation}
This suggests the following definition of the conductivity momentum,
$k_\sigma = \sigma H(t)/2$, above which the plasma conductivity destroys
any primordial spectrum. Neglecting the thermal contribution,
${\bf P}_B^{\rm th}\propto k^3$ ($k\gg k_{\sigma}$), it is worth noting that
the spectrum~(\ref{PB:spectrum:sigma}) gives
a magnetic field that is smaller than~(\ref{PB:spectrum}) on all scales.
The spectrum~(\ref{PB:spectrum}) may be understood as an upper bound and the 
spectrum~(\ref{PB:spectrum:sigma}) as a lower bound on the magnetic field
spectrum obtained by a more realistic treatment of the photon dynamics during 
radiation and matter eras.


%
%

\end{document}